\documentclass[aps,pra,showpacs,eqsecnum,twocolumn]{revtex4}
\usepackage{graphicx}
\usepackage{bm}
\begin{document}

\title{Contradictory entropic  joint uncertainty relations for complementary observables in two-level systems}

\author{Alfredo Luis}
\affiliation{Departamento de \'{O}ptica, Facultad de Ciencias
F\'{\i}sicas, Universidad Complutense, 28040 Madrid, Spain}

\begin{abstract}
We show that different entropic measures of fluctuations lead to contradictory uncertainty relations for two 
complementary observables. We apply Tsallis and  R\'{e}nyi entropies to the joint distribution emerging from 
a noisy simultaneous measurement of both observables as well as to the product of their individual statistics, 
either intrinsic or of operational origin.
\end{abstract}


\maketitle

\section{Introduction}

Historically, the joint uncertainty of pairs of observables has been mostly addressed 
in terms of the product of their variances. Nevertheless, there are situations where such 
formulation is not satisfactory enough  \cite{pvar,fase} and alternative approaches have 
been proposed, mainly in terms of diverse entropic measures \cite{CB,xx,Sl}. More specifically in 
this work we will consider the Tsallis and R\'{e}nyi entropies \cite{Ts,Ry,TRl,miTR,ycr,PP,BPP,ZZ}

In previous works we have shown that  entropic measures of fluctuations  lead to contradicting 
intrinsic joint uncertainty relations for complementary observables \cite{ycr,BPP},
among other surprising results \cite{ZZ,RL,yc}.  More specifically,  the maximum uncertainty 
states of some measures are the minimum uncertainty states of other measures, and
vice versa. By intrinsic we mean that the entropies are computed with the exact statistics 
or each observable. 

In this work we address  an operational approach where the uncertainties are 
derived from the statistics of a practical schemes devised to measure both 
observables in the same system realization. Simultaneous measurements of 
complementary observables cannot be exact and some extra uncertainty is 
unavoidable introduced \cite{jur1,jur2,njm}.  We study whether this extra uncertainty 
might affect the contradictions found in the intrinsic case. Besides, the 
simultaneous measurement provides a true joint probability distribution that 
enables an alternative assessment of joint uncertainty different from the one
provided by  the individual statistics, either intrinsic or of operational origin. 
The main conclusions we find is that the contradictions between different entropic 
measures persist, and new contradictions emerge between the entropy of the joint 
distribution and the entropies of the  product of the individual distributions.

For simplicity we  address this issue in the simplest quantum system described 
by a two-dimensional Hilbert space. This comprises very relevant practical 
situations such as the path-interference complementarity in two-beam interference.
Finally, we address further tests that might be useful to understand these contradictions.

\section{System states}

We consider two complementary observables represented by the Pauli matrices  
$\sigma_z$ and $\sigma_x$. In practical terms  they may represent path and  phase 
in two-beam interference, respectively. The system state is represented by the density 
matrix operator in the system  space  $ {\cal H}_s$ 
\begin{equation}
\rho = \frac{1}{2} \left ( I + \bm{s}  \cdot \bm{\sigma} \right ) ,
\end{equation}
where $I$ is the identity, $ \bm{\sigma}$ are the three Pauli matrices,
and  $\bm{s} = \mathrm{tr} (\rho \bm{\sigma})$ is a three-dimensional real 
vector with   $|\bm{s} | \leq 1$. For simplicity we will consider that  $\bm{s}$ lays in 
the $xz$ plane so $s_y =0$. We wil use the parametrization 
\begin{equation}
s_x = s \sin \theta, \qquad s_z = s \cos \theta ,
\end{equation}
where  $s=|\bm{s} |$. When $s=1$ the state is pure $\rho = | \psi \rangle \langle \psi |$
with 
\begin{equation}
\label{pss}
| \psi \rangle = \cos \frac{\theta}{2} | + \rangle +  \sin\frac{\theta}{2} | - \rangle ,
\end{equation}
where $|\pm \rangle$ are the eigenstates of $\sigma_z$ with eigenvalue $\pm 1$. 
Throughout we will consider the one-parameter families $S$ of pure states $s=1$
in Eq. (\ref{pss}). The exact intrinsic statistics for the $\sigma_x$, $\sigma_z$ observables 
are
\begin{equation}
\label{mis}
p_x = \frac{1}{2} \left ( 1 +  x s_x \right ),
\qquad
p_z = \frac{1}{2} \left ( 1 +  z s_z \right ) ,
\end{equation}
with $x = \pm 1$ and $z= \pm 1$.

There are two sets of states that compete to be  either the minimum  uncertainty states 
(as well as of maximum uncertainty within the set $S$) depending on the measure of 
fluctuations used \cite{ycr,BPP}.  We will  refer to them as extreme and intermediate states:

(i) The extreme states are the eigenstates of $\sigma_z$ or $\sigma_x$. These are pure 
states  in Eq. (\ref{pss}) with $\theta= m \pi/2$ for integer $m$ (these are even multiples 
$\pi/4$ and $s_x =  \pm 1, s_z=0$ or $s_x =0,  s_z = \pm 1$).  They present full certainty 
for one of the observables and complete uncertainty for the other one.

(ii) The intermediate states are the eigenstates of $\sigma_x \pm \sigma_z$. These are 
the pure states (\ref{pss}) with $\theta= (2 m +1)  \pi/4$ for integer $m$ (this is the odd 
multiples of $\pi/4$ and  $s_x = \pm s_z = \pm 1/\sqrt{2}$). They have essentially the 
same statistics for both complementary observables so they might be considered as a 
finite-dimensional counterpart of the Glauber coherent states. 

\begin{figure}
\begin{center}
\includegraphics[width=4cm]{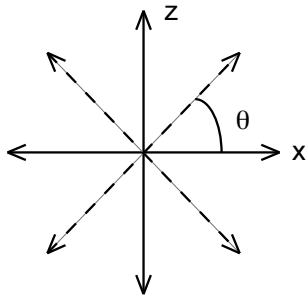}
\end{center}
\caption{Illustration of the extreme (solid line) and intermediate (dashed line) states.}
\end{figure}

In Appendix B we point out that for the intermediate states there can be no true probability
distribution underlying the operational joint distribution (\ref{js}) to be found below. This 
allows us to say that the intermediate states are literally nonclassical states regarding the 
statistics of the observables $\sigma_x$, $\sigma_z$,  while all the extreme states are fully 
classical in this respect.

\section{Joint measurement and statistics}

The  simultaneous measurement of noncommuting observables requires involving 
auxiliary degrees of freedom, usually referred to as apparatus. In our case we consider an 
apparatus described by a two-dimensional Hilbert space $ {\cal H}_a$. The measurement 
to be performed in ${\cal H}_a$  addresses the measurement of $\sigma_z$,  while  
$\sigma_x$ is measured directly on the system space ${\cal H}_s$. The system-apparatus 
coupling transferring information about $\sigma_z$ from the system to the apparatus is 
arranged via the following unitary transformation acting in $ {\cal H}_s \otimes {\cal H}_a$ 
 \begin{equation}
\label{U}
U = V_+ | + \rangle \langle + |  + V_- | - \rangle \langle - | ,
\end{equation}
where $V_\pm$ are unitary operators acting solely on ${\cal H}_a$. 

For simplicity  the initial state of the apparatus $| a  \rangle \in {\cal H}_a$ is assumed pure 
so that the system-apparatus coupling leads to  the following state (assumed pure)$ \in 
{\cal H}_s \otimes {\cal H}_a$
\begin{equation}
U | \psi \rangle | a \rangle = \cos \frac{\theta}{2} | + \rangle | a_+ \rangle +  \sin\frac{\theta}{2} 
| - \rangle  | a_- \rangle,
\end{equation}
where $| a_\pm \rangle = V_\pm | a \rangle \in {\cal H}_a$ are not orthogonal in general,
with  $\cos \delta = \langle a_+ | a_- \rangle$  assumed to be a positive real number without 
loss of generality (see Fig. 2). The measurement in ${\cal H}_a$ introducing minimum 
uncertainty is given  by projection on the orthogonal vectors $| b_\pm \rangle$  \cite{njm} (see 
Appendix A)
\begin{eqnarray}
\label{m12}
&  | b_+ \rangle = \frac{1}{\cos \phi} \left ( \cos \frac{\phi}{2}
| a_+ \rangle - \sin \frac{\phi}{2}  | a_- \rangle \right ) , & \nonumber \\ & & \\
& | b_-  \rangle =  \frac{1}{\cos \phi}  \left ( - \sin \frac{\phi}{2} | a_+ \rangle 
+ \cos \frac{\phi}{2} | a_- \rangle \right ) , & \nonumber
\end{eqnarray}
where $\phi = \pi/2 - \delta$.

\begin{figure}
\begin{center}
\includegraphics[width=4cm]{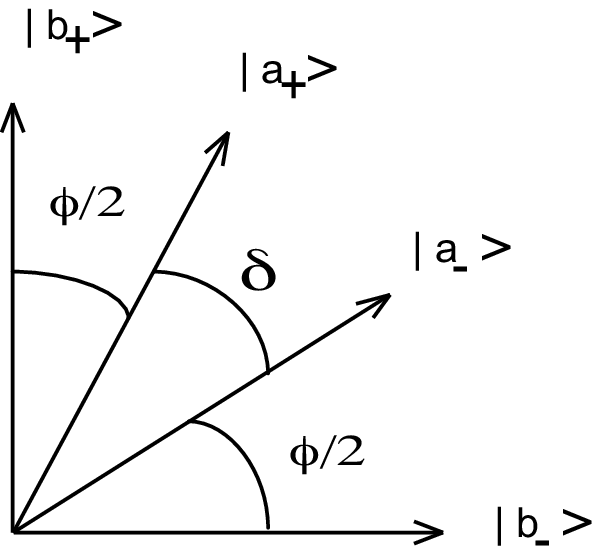}
\end{center}
\caption{Illustration of the relation between the states  $| b_\pm \rangle$ and 
$|a_\pm \rangle$.}
\end{figure}

The joint statistics for the simultaneous measurement of $\sigma_x$ in 
${\cal H}_s$ and  $| b_\pm \rangle$  in ${\cal H}_a$ is 
\begin{equation}
\label{js}
\tilde{p}_{x,z} = \frac{1}{4} \left ( 1 + z s_z \sin \delta   + x s_x \cos \delta \right  ) ,
\end{equation}
where $x=\pm1$ represent the outcomes of  the $\sigma_x$ measurement 
and $z=\pm 1$ the outcomes of the $| b_\pm \rangle$ measurement. The 
marginal statistics for both observables are 
\begin{equation}
\label{ms}
\tilde{p}_x = \frac{1}{2} \left ( 1 +  x s_x \cos \delta  \right ) ,
\qquad
\tilde{p}_z = \frac{1}{2} \left ( 1 +  z s_z \sin \delta  \right )  .
\end{equation}
When contrasted with the intrinsic statistics in Eq. (\ref{mis}) we get that the observation 
of  $\sigma_x$ is exact for $\delta= 0$ while the observation of  $\sigma_z$ is exact for 
$\phi = \delta= \pi/2$. For $\delta = \pi/4$ the extra uncertainty introduced by the unsharp 
character of the simultaneous observation is balanced between observables. 

\section{Uncertainty assessments}

We will focus on the Tsallis  $T_q$ and  R\'{e}nyi $R_q$ entropies  \cite{Ts,Ry,TRl,miTR,ZZ,PP,ycr,BPP}
\begin{equation}
T_q (p_j) = \frac{1}{1-q}  \left ( \sum_{j=1}^N p_j^q - 1\right ),
\end{equation}
and
\begin{equation}
R_q (p_j) = \frac{1}{1-q}  \ln \left ( \sum_{j=1}^N p_j^q \right ) ,
\end{equation}
where $q>0$ is a real parameter and  $p_j$ is the statistics of some observable with  $N$ outcomes. 
In both cases minimum entropy $T_q = R_q =0$  holds when all the probability is concentrated in a 
single outcome $p_j = \delta_{j,k}$ for any $k$, while maximum entropy  occurs when all the 
outcomes are equally probable $p_j= 1/N$. Both include the  Shannon entropy in the limit $q \rightarrow 1$. 
Moreover, for Gaussian-like statistics $R_q$ is proportional to variance.

The operational entropies  are always larger than the intrinsic ones $T_q ( \tilde{p}_k ) \geq T_q ( p_k ) $
and $R_q ( \tilde{p}_k ) \geq R_q ( p_k ) $ for $k=x,z$ since comparing Eqs. (\ref{ms}) and (\ref{mis}) 
we can appreciate that the observation amounts to be a reduction of $|s_k |$, and both entropies are 
monotonic functions of $|s_k |$. Their only extreme holds at the uniform distribution $p_j = 1/N$  which 
is an absolute maximum.

In order to assess the joint uncertainty of $\sigma_x$ and $\sigma_z$ using these entropies we will 
follow two strategies: 

(i)   On the one hand we will compute the Tsallis and  R\'{e}nyi entropies of  the complete joint statistics 
(\ref{js}) $T_q ( \tilde{p}_{x,z})$ and $R_q ( \tilde{p}_{x,z})$. 

(ii) On the other hand, for completeness we will consider also the entropies for the product of individual 
operational statistics $T_q ( \tilde{p}_x \tilde{p}_z )$ and $ R_q ( \tilde{p}_x \tilde{p}_z )$. In this regard 
we recall that the  R\'{e}nyi entropy is additive, 
\begin{equation}
 R_q ( \tilde{p}_x \tilde{p}_z ) =  R_q ( \tilde{p}_x ) + R_q ( \tilde{p}_z ) ,
 \end{equation}
 while in general Tsallis is not, except for $q \rightarrow 1$, \cite{Ts,Ry,miTR,PP}
 \begin{equation}
 T_q ( \tilde{p}_x \tilde{p}_z ) =  T_q ( \tilde{p}_x ) + T_q ( \tilde{p}_z ) + (1 - q ) T_q ( \tilde{p}_x ) 
 T_q ( \tilde{p}_z ) .
 \end{equation}

Both strategies will be examined and  compared with the entropy of the product of intrinsic uncertainties 
$T_q ( p_x p_z )$ and $R_q ( p_x p_z )$ for different choices of $q$ as functions of $\theta$ within the 
set $S$ of states  (\ref{pss}). For the sake of clarity and a simpler comparison  we mostly focus on normalized 
quantities of the form:
\begin{equation}
\label{norm}
T_q \rightarrow  \frac{T_q - T_{q,\mathrm{min}}}{T_{q,\mathrm{max}}- T_{q,\mathrm{min}}} ,
\end{equation}
and equivalently for $R_q$, where $T_{q,\mathrm{max}}$, $T_{q,\mathrm{min}}$ are the maximum and 
minimum of $T_q$, respectively, when $\theta$ is varied.

\section{Joint uncertainty from Tsallis and  R\'{e}nyi entropies}

\subsection{Joint uncertainty from Tsallis  entropies}

A numerical evaluation of the Tsallis entropies  $T_q ( \tilde{p}_{x,z} )$ shows that there are 
contradictions between different $q$ values. This is illustrated in Fig. 3a where we have 
represented $T_q ( \tilde{p}_{x,z} )$ for $q=1/2$ and $q=2.5$ as functions of $\theta$. It can 
be appreciated that for $q=1/2$ the minimum uncertainty states are the the intermediate states 
$\theta = \pi/4$ while for  $q=2.5$ the minimum uncertainty states are  the extreme states 
$\theta = 0, \pi/2$. 

The contradictions between different $q$ values also hold for the Tsallis entropies of the product 
statistics $T_q ( \tilde{p}_x \tilde{p}_z )$, as illustrated in Fig. 3b representing the cases $q=1/2$ 
and $q=2.5$ as functions of $\theta$. It can be appreciated that for $q=1/2$ the minimum uncertainty 
states are the extreme states $\theta = 0, \pi/2$ while for  $q=2.5$ the minimum uncertainty states 
are the intermediate states $\theta = \pi/4$. This  result coincides with the conclusions derived from 
the intrinsic entropies  $T_q (p_x p_z )$ as shown in Fig. 3c. Moreover, the case $q=1/2$ coincides with the 
conclusions of the product of variances, this is that the minimum uncertainty states are the extreme 
states. 

\begin{figure}
\begin{center}
\includegraphics[width=7cm]{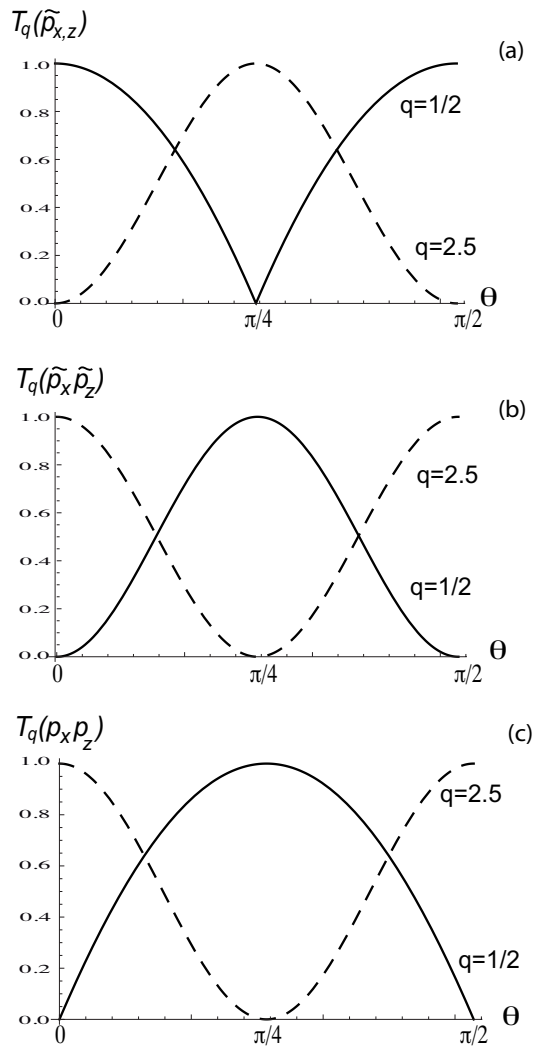}
\end{center}
\caption{(a) Plot of the Tsallis entropies for the joint statistics $T_q (\tilde{p}_{x,z} )$ for $q=1/2$ (solid line) 
and $q=2.5$ (dashed line) as functions of $\theta$. (b) Same as before but for the Tsallis entropies of the 
product of operational statistics $T_q (\tilde{p}_x \tilde{p}_z )$. (c) Same as before but for the Tsallis 
entropies of the product of intrinsic statistics $T_q (\tilde{p}_x \tilde{p}_z )$. All quantities have been 
properly normalized to lay between 0 and 1 as in Eq. (\ref{norm}).}
\end{figure}

In Fig. 4 we have represented the extreme-intermediate competition for minimum uncertainty 
by plotting the difference of their Tsallis entropies or different values of $q$
\begin{equation}
\label{iec}
\delta T_q  (\tilde{p}_{x,z} )  \propto \left . T_q (\tilde{p}_{x,z} ) \right |_{\theta = \pi/4}  -
 \left . T_q ( \tilde{p}_{x,z} ) \right |_{\theta = 0}  .
\end{equation}
When $\delta \tilde{T}  >0$ the extreme stats are of minimum uncertainty while when  $\delta \tilde{T}  
< 0$ the intermediate states are the minimum uncertainty states. We have also included the same quantity 
computed for  the Tsallis entropies of the product of operational $\delta T_q (\tilde{p}_x \tilde{p}_z  )$ and
intrinsic $\delta T_q (p_x p_z )$ statistics. 

We can appreciate that the contradictions between different $q$ values  for $T_q (\tilde{p}_{x,z} )$ 
take place between entropies with $q \in (2,3)$ and $q \in (0,2) \cup (3,\infty )$. For the intrinsic case
$T_q (p_x p_z  ) $ contradictions arise between the $q$ values $q<1.4313$ and $q>1.4313$, while for
$T_q (\tilde{p}_x \tilde{p}_z  ) $ they hold between  $q<1.3439$ and $q>1.3439$.  Thus, observation 
drives the critical $q$-values with respect to the intrinsic case.  Moreover, there are two $q$ intervals
 $q \in (1.4313, 2)   \cup  (3, \infty )$ free of  these contradictions.

\begin{figure}
\begin{center}
\includegraphics[width=7cm]{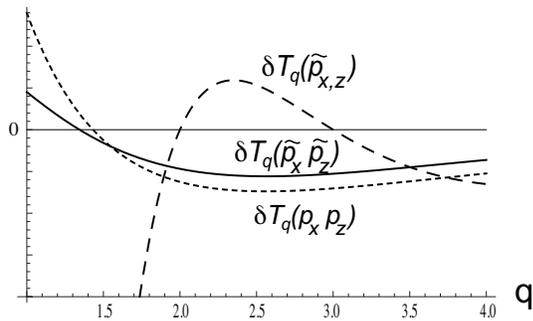}
\end{center}
\caption{Difference between the uncertainty for intermediate and extreme states  for the entropies 
$\delta T_q  (\tilde{p}_x  \tilde{p}_z  )$ in Eq. (\ref{iec})  (solid line),  and equivalent quantities 
for the entropy of the joint statistics  $\delta T_q (\tilde{p}_{x,z} ) $ (dashed line), and the ntrinsic 
 entropies $\delta T_q (p_x p_z ) $ (dotted line), as functions of $q$. Arbitrary units have been used 
 different for each curve to clearly display the positive and  negative regions.}
\end{figure}

These two figures  clearly shows the double kind of contradictions we have found: (i) We have different 
conclusions for different $q$ values, both in the intrinsic and operations frameworks. (ii) For the same 
$q$ we have contradictions between the entropy of the joint distribution and the entropy of the 
product of individual distributions, irrespectively of whether this is  intrinsic or operational. 

\subsection{Joint uncertainty from R\'{e}nyi entropy}

A numerical evaluation of the R\'{e}nyi entropies reproduces the contradictions between different $q$ 
values identical to the ones represented in Fig. 3.  Also R\'{e}nyi entropies reproduce the extreme-intermediate 
competition for minimum uncertainty for different values of $q$ illustrated in Fig. 4, with identical critical $q$ 
values. 

\section{Further tests}

In this section we compute entropic-related quantities that might serve to test the meaning of the 
contradictions revealed in the preceding section.

\subsection{Difference of entropies}

The \textit{bona fide} joint probability distribution given by the simultaneous measurement 
provides us a meaningful comparison of the entropies for the joint $\tilde{p}_{x,z}$ and product 
$\tilde{p}_x \tilde{p}_z$ statistics  that may be helpful to understand the contradicting behaviors 
reported above by providing new information about these quantities. 

For example we can compute the differences 
\begin{equation}
\label{ed}
T_q ( \tilde{p}_x \tilde{p}_z ) - T_q ( \tilde{p}_{x,z})  ,  \qquad
R_q ( \tilde{p}_x \tilde{p}_z ) - R_q ( \tilde{p}_{x,z}) ,
\end{equation}
where we have that
\begin{equation}
\tilde{p}_{x,z}- \tilde{p}_x \tilde{p}_z = - \frac{1}{16} x z s^2 \sin (2 \theta) \sin (2 \delta) .
\end{equation}
The difference of the statistics is maximum for the simultaneous measurement that treat forth observables 
symmetrically $\delta = \pi /4$ and for intermediate states $ \theta = \pi /4$.

The Tsallis entropy differences are plotted in Fig. 5a showing that for certain $q$ values we 
have the paradoxical result that the entropy of the joint distribution $\tilde{p}_{x,z}$ is larger 
than the entropy of the product of the marginal distributions $\tilde{p}_x \tilde{p}_z$. In Fig. 5b
we have plotted the difference of entropies for intermediate states revealing the $q$ values 
where the counterintuitive result  $T_q ( \tilde{p}_x \tilde{p}_z ) < T_q ( \tilde{p}_{x,z})$ holds, 
this is $q > 1.60$. Note that this counterintuitive result cannot account for the contradictions 
reported above since the corresponding  $q$ regions are different. 

 Here again, with the R\'{e}nyi entropies  we get the same result with the same critical value
 $q=1.60$. Differences between $R_q (\tilde{p}_{x,z})$ and $R_q (\tilde{p}_x \tilde{p}_z)$ 
regarding minimum uncertainty states have been also found in Ref. \cite{jur2} .

\begin{figure}
\begin{center}
\includegraphics[width=7cm]{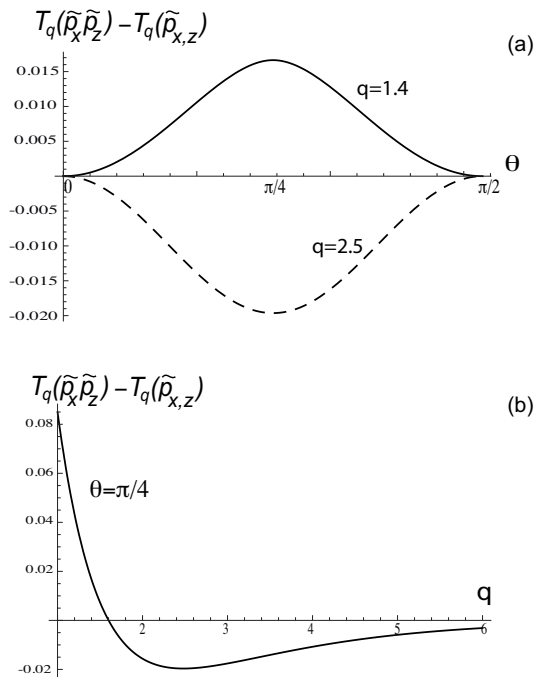}
\end{center}
\caption{(a) Difference of Tsallis entropies $T_q ( \tilde{p}_x \tilde{p}_z ) - T_q ( \tilde{p}_{x,z})$ as functions
of $\theta$ for $q=1.4$ (solid line) and $q=2.5$ (dashed line). (b) Same difference of Tsallis entropies 
as function of $q$ for intermediate states $\theta = \pi/4$ showing counterintuitive behavior for $q >1.60$. }
\end{figure}

 \subsection{Mutual information}
 
 For Shannon entropy there is an equivalence between mutual information and the entropy differences 
 (\ref{ed}). For Tsallis and  R\'{e}nyi entropies such equivalence no longer holds in general. Thus it may be 
 worth investigating whether mutual information displays any unexpected behavior.  To this end we have 
 examined the following definitions of Tsallis and R\'{e}nyi mutual informations \cite{miTR}
 \begin{eqnarray}
 & I_{T,q }= \frac{1}{q-1} \left [ \sum_{x,z} \frac{\tilde{p}_{x,z}^q}{\left ( \tilde{p}_x \tilde{p}_z \right )^{q-1} } - 1 \right ] , 
 & \nonumber \\
&  I_{R,q} = \frac{1}{q-1} \ln \left [  \sum_{x,z} \frac{\tilde{p}_{x,z}^q}{\left ( \tilde{p}_x \tilde{p}_z \right )^{q-1} } \right ] . &
  \end{eqnarray}
We have found no contradiction nor counterintuitive behavior since we have obtained always 
  $I_{T,q } \geq 0$ and   $I_{T,q } (\mathrm{intermediate} ) \geq I_{T,q } ( \mathrm{extreme})$, and equivalently 
  for R\'{e}nyi entropies. 
   
\subsection{Generalized Fisher information}

As a further test of practical origin we may consider the performance provided by the states in Eq. (\ref{pss})
in the detection of  signals encoded by transformations generated by $\sigma_y$ in the vicinity of the identity. 
Considering the measurement of $\sigma_z$ after the transformation, the metrological performance  can be 
given in terms similarity between the transformed and original $\sigma_z$-statistics. The most frequent test 
is given by Fisher information. Nevertheless, and following the same spirit of Tsallis and R\'{e}nyi entropies, 
several $q$-order generalizations of Fisher information have been provided in Ref. \cite{RL} in the form 
\begin{equation}
\label{Fq}
F^q_q =\left [ \sum_{j=1}^N p_j  \left | \frac{d}{d \eta} \ln p_j \right |^{1/q} \right ]^q ,
\end{equation}
where $p_j$ is the measured statistics depending on the parameter $\eta$ of the transformation, and the 
final expression is to be evaluated at $\eta =0$. In our case there are only two outcomes with 
probabilities $p_+ = \cos^2 ( \eta - \theta/2)$ and $p_- = \sin^2 ( \eta - \theta/2)$.

In Fig. 6 we have evaluated $F^q_q$ finding again contradictions between $q>1/2$ (intermediate states 
providing maximum  $F^q_q$) and  $q <1/2$ (intermediate states providing minimum  $F^q_q$), while 
all states provide the same  $F^q_q$ in the case iof the standard Fisher information $q=1/2$.
These plots are similar to the ones obtained in Ref. \cite{RL}
  
\begin{figure}
\begin{center}
\includegraphics[width=7cm]{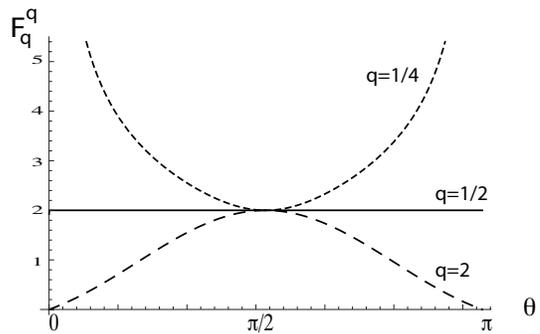}
\end{center}
\caption{Plot of the generalized Fisher information $F^q_q$ as function of $\theta$ 
for $q=1/2$ (solid line), $q=2$ (dashed line), and $q=1/4$ (dotted line) .}
\end{figure}

\subsection{Extension to unbounded continuous Cartesian variables}

For the sake of comparison we extend the analysis to a pair of complementary unbounded continuous 
Cartesian variables, such as one-dimensional position and linear momentum of a particle, or the 
quadratures of a single-mode electromagnetic field. Let us focus on the quadrature case with a pair 
of quadrature operators  $X$, $Y$,  satisfying the commutation relation $ [X,Y] = 2 i$. Furthermore, let 
us assume 
pure states with Gaussian intrinsic statistics 
\begin{equation}
p_k = \frac{1}{\sqrt{2 \pi} \Delta k} \exp \left [ - \frac{k^2}{2 (\Delta k)^2} \right ], 
\end{equation}
for $k=x,y$ where $\Delta k$ are the corresponding variances and we have assumed without loss 
of generality that $\langle k \rangle =0$. We focus on the set $S$ of pure, 
minimum uncertainty states for the product of variances  $\Delta x \Delta y = 1$. This set includes 
the analog of the intermediate states when $\Delta x = \Delta y =1$, which are the Glauber 
classical-like coherent states. The set $S$ also includes the extreme states as the limits 
 $\Delta x \rightarrow 0$ and $\Delta y \rightarrow 0$  (which are nonclassical arbitrarily squeezed states 
 tending to be eigenstates of a quadrature operator). 
 
We consider a version of Tsallis  entropy adapted to continuous distributions as
\begin{equation}
T_q (p_k) = \frac{1}{1-q}  \left ( \int_{-\infty}^\infty  dk \; p_k^q - 1 \right ),
\end{equation}
for $k=x,y$, and similarly for the  R\'{e}nyi entropies. It is worth noting that in this case entropies are no longer 
positive definite.

In the first place we can note that within the set $S$ there are no contradictions between different $q$ 
 measures for the Tsallis and R\'{e}nyi entropies applied to the product of intrinsic statistics
\begin{equation}
 T_q ( p_x  p_y ) = \frac{1}{1-q} \left [ \frac{\left ( 2 \pi \Delta x \Delta y \right )^{1-q}}{q} - 1 \right ] ,
 \end{equation}
that take the same value for all the states with  $\Delta x \Delta y = 1$. Similar conclusion is obtained 
for the R\'{e}nyi entropies
 \begin{equation}
 R_q (p_x  p_y ) = \ln \left ( \Delta x \Delta y \right ) + \ln (2 \pi) - \frac{1}{1-q} \ln q .
 \end{equation}

 Regarding the simultaneous joint measurement of $X$ and $Y$ we resort to standard schemes \cite{njm}
 leading to a joint distribution 
 \begin{equation}
 \tilde{p}_{x,y} = \frac{\exp \left \{ - \frac{x^2}{2 \left [ 1 + (\Delta x )^2 \right ] } - 
 \frac{y^2}{2 \left [ 1 + (\Delta y )^2 \right ] } \right \} }
 {2 \pi \sqrt{ \left [ 1 + (\Delta x )^2 \right ]  \left [ 1 + (\Delta y )^2 \right ] }} .
 \end{equation}
 Let us note that in this case $\tilde{p}_{x,y} = \tilde{p}_x \tilde{p}_y$ so that there is no difference
 between the joint statistics and its product of marginals. The operational uncertainty is now
 \begin{equation}
 T_q ( \tilde{p}_{x,y} ) = \frac{1}{1-q} \left \{ \frac{\left [ 2 \pi \sqrt{ \left ( 1+\Delta x^2 \right )
 \left (1+\Delta y^2 \right ) } \right ]^{1-q}}{q} - 1 \right \} .
 \end{equation}
For all the $q$ values that we have considered we have found no contradictions between 
different $q$ values within the set of states  $\Delta x \Delta y = 1$, so that the coherent states
 $\Delta x = \Delta y = 1$ are always the minimum uncertainty states. The same conclusion is 
 obtained using  R\'{e}nyi entropies. Nevertheless this does not exclude that contradictions 
 might be found for other state families.
  
 Nevertheless, it might be worth noting that we have found contradictions between different 
 $q$ values when considering the sum of Tsallis entropies for both observables $\sigma_x$ 
 and $\sigma_z$,  both intrinsic and operational, i. e.,
  \begin{equation}
 \label{Cic}
 T_q ( p_x ) + T_q ( p_y ) = \frac{(2\pi)^{\frac{1-q}{2}}}{\sqrt{q} (1-q)} \left [ \left ( \Delta x \right )^{1-q}
 + \left ( \Delta y \right )^{1-q}  - 2 \right ] ,
 \end{equation}
 and
 \begin{eqnarray}
 \label{Coc}
 & T_q ( \tilde{p}_x ) + T_q ( \tilde{p}_y ) = \frac{(2\pi)^{   \frac{1-q}{2}}}{\sqrt{q} (1-q)} & \nonumber \\
 & \times \left \{ 
 \left [ 1 + (\Delta x )^2 \right ]^{\frac{1-q}{2}} +   \left [ 1 + (\Delta y )^2 \right ]^{\frac{1-q}{2}}  - 2 .
  \right \} . &
 \end{eqnarray}
 We have plotted both in Fig 7 as a function of $\Delta x$ for different values of $q$ showing that 
 the coherent states $\Delta x =1$ can be either maximum or minimum.  The transition from maximum 
 to minimum is dictated by the $q$ value for which vanish the second derivative the  left-hand sides 
 of Eqs. (\ref{Cic}) and (\ref{Coc}) with  respect to $\Delta x$ evaluated at $\Delta x = 1$. In the intrinsic case
 this  holds for $q=1$, this is for the Shannon entropy, while in the operational case it holds for $q=3$. 
 Despite we are aware of the lack of additivity of Tsallis entropies we find this result still striking. To some 
 extent one might regard the difference between $T_q (p_j p_k)$ and $T_q (p_j)+T_q( p_k)$ as merely 
 quantitative, but the qualitative difference pointed out by the above example might add valuable arguments 
 to the discussion of whether $T_q (p_j)$ is a meaningful assessment of quantum fluctuations.
  
\begin{figure}
\begin{center}
\includegraphics[width=7cm]{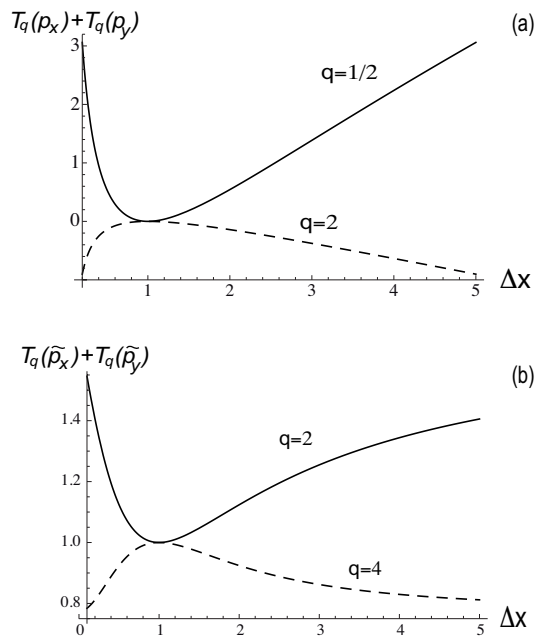}
\end{center}
\caption{(a) Plot of  $ T_q ( p_x ) + T_q ( p_y ) $ as a function of $\Delta x$ for $q=1/2$ (solid line) 
and $q=2$ (dashed line). (b) Plot of  $ T_q ( \tilde{p}_x ) + T_q ( \tilde{p}_y ) $ as a function of 
$\Delta x$ for $q=2$ (solid line) and $q=4$ (dashed line) and normalized to their value at $\Delta x =1$.  }
\end{figure}

\section{Discussion}

We haver presented several examples of application of Tsallis and R\'{e}nyi entropies as potential
measures of quantum uncertainty. To this end we are considering situations where quantum features 
are relevant, such as uncertainty relations and metrology. We hope that the contradicting cases 
presented in this work might proper grounds for the definition of suitable measures of quantum 
uncertainty beyond variance. Besides, they might provide different interpretations of the general 
idea of quantum uncertainty and provide different measures adapted to these new situations.

\section*{ACKNOWLEDGMENTS}

A. L. acknowledges support from Project No. FIS2012-35583
of the Spanish Direcci\'{o}n General de Investigaci\'{o}n del 
Ministerio de Econom\'{\i}a y Competitividad, and from Project 
QUITEMAD S2009-ESP-1594 of the Consejer\'{\i}a de Educaci\'{o}n
de la Comunidad de Madrid.

\appendix

\section{Apparatus measurement introducing minimum uncertainty}

Here we examine the optimum measuring strategy introducing a minimum additional noise in the 
joint measurement. On the one hand, the operational statistics $\tilde{p}_x$ does not depend on 
the measurement to be performed on the apparatus. This is because it corresponds to the 
measurement of $\sigma_x$ in the reduced state obtained after tracing over the apparatus degrees 
of freedom. On the other hand, the operational statistics $\tilde{p}_z$  can be described by projection 
on two orthogonal states $|b_\pm \rangle$ in the subspace spanned by $|a_\pm \rangle$ on the apparatus 
space $\mathcal{H}_a$, as illustrated in Fig. 8. 

\begin{figure}
\begin{center}
\includegraphics[width=5cm]{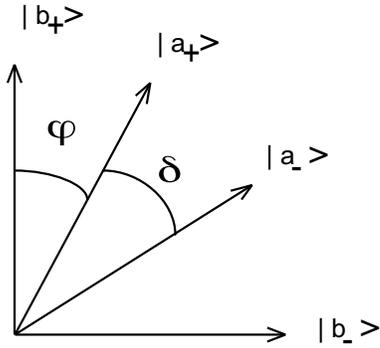}
\end{center}
\caption{Illustration of the relation between the states  $| b_\pm \rangle$ and $|a_\pm \rangle$.}
\end{figure}

The objective is to look for the choice of angle $\varphi$ leading to minimum uncertainty.
The observed $\tilde{p}_z$ and intrinsic  $p_z$ probabilities are related in the form
\begin{equation}
\pmatrix{\tilde{p}_+ \cr \tilde{p}_-  } = M \pmatrix{p_+ \cr p_-  } , \quad
M = \pmatrix{\cos^2 \varphi  & \cos^2 (\varphi + \delta ) \cr \sin^2 \varphi  & \sin^2 (\varphi + \delta ) }  .
\end{equation}
The condition of minimum uncertainty can be implemented as the minimum distance between 
$\tilde{p}_z$ and  $p_z$, this is the minimum distance between $M$ and the identity matrix $I$. 
For example we may look for the minimum of 
 \begin{equation}
 \textrm{tr} \left [ \left ( M- I \right ) ^2 \right ] = \left [ \sin^2 \varphi + \cos^2 \left ( \varphi + \delta \right ) \right ]^2 ,
\end{equation}
which is obtained for $2 \varphi = \pi/2 - \delta$. 

 \section{Nonclassicality after joint statistics}

Since the measurement scheme is known it is possible to formally invert the process to obtain 
from the joint unsharp statistics $\tilde{p}_{x,z}$ in Eq. (\ref{js}) a sharp one $p_{x,z}$
providing the correct marginals\cite{WMM}. 

Regarding the intrinsic $p_k$ and operational $\tilde{p}_k$ statistics as two-dimensional column 
vectors, from Eq. (\ref{ms})  we have $\tilde{p}_k =M_k p_k$ with
\begin{equation}
M_k = \pmatrix{ \cos^2 \phi_k & \sin^2 \phi_k \cr \sin^2 \phi_k & \cos^2 \phi_k},
\end{equation}
with  $k=x,z$, being $\phi_x = \delta/2$, and $\phi_z = \pi /4 - \delta /2$. 

The relation  $\tilde{p}_k =M_k p_k$ can be inverted to obtain the intrinsic "true" or noiseless 
distribution in terms of the operational one  $p_k = M_k^{-1} \tilde{p}_k$ . Thus, regarding the 
operational joint distribution $\tilde{p}_{x,z}$ as a $2 \times 2$ matrix we may infer a "true" or 
"noiseless" joint distribution $p_{x,z}$ as  
\begin{equation}
p_{x,z} = M_x^{-1} \tilde{p}_{x,z} M^{-1,t}_z = \frac{1}{4} \left ( 1 + z s_z   + x s_x  \right  ) ,
\end{equation}
where the superscript $t$ denotes matrix transposition. This distribution $p_{x,z} $ provides the 
true intrinsic distributions $p_x$ and $p_z$ as its proper marginals.

Quantum mechanics in this case is reflected in the fact that for some states $p_{x,z}$ takes 
negative values. These states can be termed nonclassical. Concerning our problem we note 
that the extreme states are classical  $p_{x,z} \geq 0$ (say for example $p_{x,z}=(1+z)/4$ for 
$s_x=0$, and $s_z=1$), while the intermediate states are all nonclassical  $p_{x,z} < 0$, since 
for example for $s_x = s_z = 1/\sqrt{2}$ we get $p_{-1,-1} = (1-\sqrt{2})/4 < 0$, and similarly for 
all the other intermediate states. Let us note that this is precisely the opposite behavior we find 
for the unbounded Cartesian variables, where the classical states are the intermediate states 
(the Glauber coherent states), while the nonclassical states are the extreme states (limit of arbitrarily 
squeezed states).

\end{document}